\documentclass[12pt,letterpaper,fleqn]{article}
\usepackage{amsthm,amsfonts,amsmath}
\usepackage{epsfig}
\usepackage{graphics}
\usepackage{subfigure}

\newcommand{\be}{\begin{enumerate}}
\newcommand{\ee}{\end{enumerate}}
\newtheorem{thm}{Theorem}[section]

\newtheorem{prop}[thm]{Proposition}
\newtheorem{cor}[thm]{Corollary}
\newtheorem{lem}[thm]{Lemma}

\newtheorem{rem}[thm]{Remark}
\newtheorem{ex}[thm]{Example}

\begin{document}

\title{The Economic Default Time and the Arcsine Law}
\author{ Xin Guo \thanks{
Dept of Industrial Engineering and Operations Research, UC at Berkeley, Berkeley, CA 94720-1777.
Email: xinguo@ieor.berkeley.edu. Tel: 1-510-642-3615.} \ \ \ \ Robert A.
Jarrow \thanks{%
Johnson Graduate School of Management, Cornell University, Ithaca, New York
14853, USA and Kamkura Corporation. email: raj15@cornell.edu. Tel: 1-607-255-4729. }
\ \ \ \ Adrien de Larrard\thanks{
175, rue du Chevaleret, Laboratoire de Probabilit\'es et Mod\`eles Al\'eatoires,
75013, Paris, France. Email: larrard@clipper.ens.fr. Tel: 336-69931380} }
\date{\today}
\maketitle

\begin{abstract}
This paper develops a structural credit risk model to characterize the
difference between the economic and recorded default times for a firm.
Recorded default occurs when default is recorded in the legal system. The
economic default time is the last time when the firm is able to pay off its debt prior to the legal default time. It has been empirically documented that these two times
are distinct (see Guo, Jarrow, and Lin (2008)). In our model, the
probability distribution for the time span between economic and recorded
defaults follows a mixture of Arcsine Laws, which is consistent with the
results contained in Guo, Jarrow, and Lin. In addition, we show that the
classical structural model is a limiting case of our model as the time
period between debt repayment dates goes to zero. As a corollary, we show
how the firm value process's parameters can be estimated using the tail
index and correlation structure of the firm's return.
\end{abstract}

\bigskip

\bigskip

\bigskip

\bigskip

\bigskip

\section{Introduction}

The credit risk crisis of 2007 and the resultant financial institution
failures, including Lehman Brothers in September of 2008, lead to the
Dodd-Frank financial reform act being passed by the U.S. Congress in 2010.%
\footnote{%
See New York Times, May 20, 2010, "Bill Passed in Senate Broadly Expands
Oversight of Wall St.", David Herszenhorn.} This regulatory reform act
emphasizes the importance of financial institutions retaining sufficient
equity capital to avoid future failures. Essential in the determination of a
financial institution's equity capital is an accurate assessment of the
risks of the institution's investments. The majority of these investments
have credit or counterparty default risk. A notable example, and a key
culprit in the credit crisis, were the bonds issued by collateralized debt
obligations (CDOs). To calculate the risk of the institution's investments,
estimating the loss distribution is essential. Crucial in this loss
distribution is modeling the default processes for the relevant credit
entities. It is well known that the default process can be completely
characterized by a random default time and a loss given default or
equivalently, a recovery rate process. The estimation of this default
process has received significant study in the academic literature (see, for
example, Chava and Jarrow (2004), Campbell, Hilscher, Szilagyi (2006), and
Bharath and Shumway (2008), and Chava, Stefanescu and Turnbull (2006)).

To obtain accurate estimates of the default process, the timing of default
needs to be correctly measured. A recent empirical study by Guo, Jarrow, and
Lin (2008) on the time-series behavior of market debt prices around the
recorded default date reveals a surprising fact. By studying approximately
20 million price quotes and execution prices from almost 31 thousand
different bond issues between Dec 21, 2000 and October 2007, they observed
that for the majority of the defaulted firms, the market anticipates the
default event well before default is recorded. To emphasize this
distinction, they defined the \textquotedblleft economic default
date\textquotedblright\ as the first date the market prices the debt as if
it has defaulted.

\begin{figure}[tbh]
\centering
\includegraphics[width=7cm]{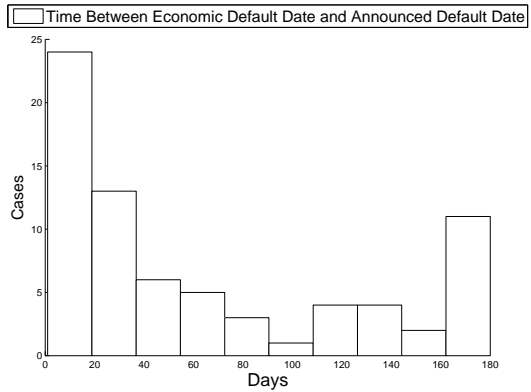}
\caption{Time between the economic and recorded default dates.}
\label{DIFF1}
\end{figure}

Guo, Jarrow, and Lin's statistical analysis shows that the time span between
the economic and recorded default dates has a significant impact on recovery
rate estimates and is key to obtaining unbiased estimates for defaultable
bond prices. To illustrate this difference, consider Figure \ref{DIFF1}.
This figure contains a histogram of the time span between the economic and
recorded default dates. Of the 96 debt issues studied, 73 trigger economic
default strictly before the recorded default date. This difference is
significant. Of these 73 issues, 13 have the economic default date at least
180 days before the recorded default date. Capponi (2009) confirms, by a
subsequent and different statistical test using credit default swaps (CDS)
and equity prices, the market's pre-knowledge of the onset of default.

\begin{table}[tbp] \centering%
%
%
%
%
\end{table}%

To facilitate future estimation of default times and recovery rates, the
purpose of this paper is to develop a theoretical model for a firm's default
process that embeds a distinction between an economic and a recorded default
time. To our knowledge, ours is the first paper to provide such a model.

The existing literature on modeling credit risk is extensive, see Jarrow
(2010) for a recent review. There are two types of credit risk models:
structural and reduced-form. These are related via the information sets
available to the management and the market. Reduced form models can be
obtained from structural models, by conditioning on the smaller information
sets available to the market. In this sense, structural models are the more
general approach, from which the reduced form models can be generated.
Because we are interested in constructing a new model that includes a
distinction between the economic and recorded default dates, we adopt the
structural modeling approach. Extensions of our model considering reduced
information sets is left for subsequent research.

The initial class of structural models (see, for example, Merton (1974),
Black and Cox (1976), Longstaff and Schwartz (1976)) had default occurring
when the firm's asset value hits an exogenously given default barrier.
Leland (1994) and Leland and Toft (1996), among others, extended this class
of structural models by endogenizing the default barrier. In this extension,
the default barrier was selected to maximize the firm's equity value. For
analytical reasons, this extended class of models assumes that the firm's
debt is perpetual and that both the changes in the magnitude and repayment
of the firm's debt is continuous. An implication of this structure is that
the firm's default time can occur at any instant continuously in time.

In contrast, and more consistent with market practice with respect to the
repayment of a firm's debt (both principal and interest), we assume that
\textit{the firm only makes required debt payments on a set of predetermined
discrete times}. This simple modification enables us to introduce a
distinction between the firm's economic and recorded default times. In this
regard, the recorded default time $\tau _{r}$ is defined to be the first
such \textit{discrete} time that a required debt payment is missed. In
contrast, the economic default time $\tau _{e}$ is defined to be the last
time prior to $\tau _{r}$ that the firm is able to make such a required debt
payment. 
Since the firm's management is aware of the fact that this is the last such
debt repayment, the economic default time represents the \textquotedblleft
true\textquotedblright\ time of default.

Herein, we study the probability distributions of the economic and recorded
default times, $\tau _{e}$ and $\tau _{r}$,\ respectively. The main analytic
difficulty is that $\tau _{e}$ is not a first passage time, but a last
passage time given the distribution of $\tau _{r}$. The key and simple idea
underlying our solution is to decompose $\tau _{e}$ into $\tau _{r}-\tau
_{e} $ and $\tau _{r}$. Our primary analytical tool is the fluctuation
theory of random walks. The analysis also uses the extensive literature on L%
\'{e}vy processes (see Bertoin (1996) and the references therein), in
particular on the links between the law of overshoot for the first passage
time and the renewal measures of a L\'{e}vy process by Doney and Kyprianou
(2006). By studying the distribution of $\tau _{r}-\tau _{e}$, the time
difference between recorded and economic default times, we show that it is a
mixture of the \textit{arcsine law}. This result is (surprisingly)
consistent with the empirical observation by Guo, Jarrow, and Lin (2008)
(Figure \ref{DIFF1}).

In addition, we show that the classical structural model can be obtained as
a limiting case of our model as the time between the debt repayment dates
goes to zero. Our approach illustrates how to use the tail index and
correlation structure of the firm's return to estimate the value process's
parameters (either Brownian motion based or more general L\'{e}vy models).

An outline of this paper is as follows. Section 2 presents the model.
Section 3 characterizes the distributions for the economic and recorded
default times, section 4 relates our model to the classical structural
model, and section 5 concludes.

\section{The Model}

We start with a filtered probability space $(\Omega ,\mathcal{F},\mathcal{F}%
_{t},P)$ that satisfies the usual conditions. Traded are a firm's asset
value process $S_{t}$ and a money market account. Without loss of
generality, we assume that the spot rate of interest is zero, so that the
money market account has a constant value across time (unity).

\subsection{The Firm Value Process}

Given a filtered probability space $(\Omega ,\mathcal{F},\mathcal{F}_{t},P)$
that satisfies the usual conditions, the value of the firm $S=(S_{t})_{t\ge
0} $ follows a geometric L\'{e}vy process together with its natural
filtration $({\mathcal{F}}_t)_{t\ge 0}$. Specifically, $S_{t}$ is an
exponential of a $(b,\sigma ,\nu )$ L\'{e}vy process, i.e.
\begin{equation}  \label{eqn:model}
S_{t}=S_{0}\exp (X_{t}),
\end{equation}%
where%
\begin{equation}
E[e^{i\theta X_{t}}]=e^{t\psi (\theta )},
\end{equation}%
with
\begin{equation*}
\psi (\theta )=ib\theta -\frac{\sigma ^{2}}{2}\theta ^{2}+\int_{(-\infty
,\infty )}(e^{i\theta x}-1-i\theta x1_{|x|\leq 1})\nu (dx)+\int_{(-\infty
,\infty )}(e^{i\theta x}-1)1_{|x|\leq 1})\nu (dx),
\end{equation*}%
and the jump measure $\nu $ satisfying
\begin{equation*}
\int_{(-\infty ,\infty )}(1\wedge x^{2})\nu (dx)<\infty .
\end{equation*}
Note that $(S_{t},t\geq 0)$ is a martingale if the following condition is
satisfied:
\begin{equation*}
b+\dfrac{\sigma ^{2}}{2}+\int_{\mathbb{R}}(e^{z}-1-z1_{z\leq 1})\nu (dz)=0.
\end{equation*}%
Consistent with $P$ being an equivalent martingale measure, we assume that
this condition is satisfied.  This martingale
measure need not be unique, so the market need not be complete.

\subsection{The Debt Structure}

Before introducing the definition of default times, we first need to
characterize the firm's debt structure. 
We assume that \textit{the firm needs to make debt
repayments at a predetermined (deterministic) set of discrete times}, denoted%
\textit{\ $N_{1},N_{2},\ldots ,N_{n}$}.

For simplicity, let $N_{k}=kN$ for a fixed $N>0$, which is the time between
two consecutive debt repayments (for instance, quarterly). At time $N_{k}$,
the amount of debt in the firm is $D_{k}$. For simplicity, we assume that $%
D_{k} = D$ is constant over time.

Consistent with a structural model, the \textit{recorded default time} $\tau
_{r}$ is the \textit{first time} the firm is unable to make a debt
repayment, i.e.%
\begin{equation}
\tau _{r}=\inf \{N_{k},S_{N_{k}}\leq D\}.
\end{equation}%
As long as the firm's asset value exceeds the barrier, it can liquidate
assets as necessary to make the debt repayment. Below the barrier, however,
it does not have sufficient resources to fulfill its debt obligation.

We define the \textit{economic default time} to be the \textit{last time},
before the onset of recorded default, when the firm is able to make a debt
repayment, i.e.,
\begin{equation}
\tau _{e}=\sup_{\tau _{r} \geq t \geq \tau _{r}-N}\{t,S_{t}\geq D\}.
\end{equation}%
Since the firm's management is aware of the fact that this is the last such
debt repayment time, the economic default time represents the firm's
``true'' default time. In contrast, the recorded default time only
corresponds to when this event becomes "official" and is recorded in both
the legal system and the default databases.

The main difference between our model and the classical structural models
comes from the discrete debt repayment times. For the existing models,
recorded default can happen at any time. In our model, recorded default can
only happen at\ one of the discrete times $N_{k}$. This discreteness enables
us to distinguish between the economic and recorded default times as
discussed above.

\begin{rem}
In Guo, Jarrow, and Lin (2008), the economic default is defined using the
entire trajectory of a firm's bond price until a recorded default occurs in
the database. Once recorded default has occurred, one goes backward in time
until the discounted price of the bond is the same as the price of an
otherwise equivalent defaulted bond. In this paper we provide an economic
model for the recorded default time and we define the economic default time
using the path of the firm's value process. Our definition of economic
default is consistent with that contained in Guo, Jarrow, and Lin (2008).
\end{rem}

\section{Computing $\protect\tau _{r}$ and $\protect\tau _{e}$}

This section computes the distributions for the recorded default time $\tau
_{r}$, the economic default time $\tau _{e}$, and relates them to the
Arcsine Law. First, note that $\tau _{e}$ is not a stopping time, but rather
a last passage time conditioned on the knowledge of $\tau _{r}$. Direct
computation of $\tau _{e}$ seems complicated. To facilitate the computation,
we compute $\tau _{e}$ in two steps. First, we we consider $\tau _{r}-\tau
_{e}$. Second, we compute $\tau _{r}$, which is the first passage time of a
random walk. The distribution of $\tau _{e}$ follows naturally from these
two steps.

\subsection{The Arcsine Law and $\protect\tau _{r}-\protect\tau _{e}$}

In this section, we show that the distribution of $\tau _{r}-\tau _{e}$ is a
mixture of Arcsine laws.

To start, let us fix some notation. First, we define $(u_{n}(x),n\geq 0)$ to
be a sequence of recorded default probabilities. That is, for all $n\geq 0$,
\begin{equation}
u_{n}(x)=\mathbb{P}[\tau _{r}=nN|S_{0}=x]=\mathbb{P}_{x}[\tau _{r}=nN],
\end{equation}%
where $\mathbb{P}_{x}$ is the probability measure of the process $(S_{t}, t
\geq 0)$ conditioned on $S_{0} = x$.

Note that
\begin{equation*}
\mathbb{P}_{x}[\tau_{r} < \infty]=\sum_{n=1}^{\infty }u_{n}(x)
\end{equation*}
and for any $0\le p<k\le K$,
\begin{equation*}
\mathbb{P}_{x}[pN\leq \tau _{r}\leq kN]=\sum_{n=p}^{k}u_{n}(x).
\end{equation*}

Next, define for $n\geq 0$ the time between real default and economic
default conditioned on defaulting at time $nN$,
\begin{equation}
\phi _{n}(x,s)=\mathbb{P}_{x}[\tau _{r}-\tau _{e}\in ds|\tau _{r}=nN].
\end{equation}%
In particular, for $x\geq D$, the probability that default occurs at $t=N$
is
\begin{equation}
\phi (x,s)=\mathbb{P}_{x}[\tau _{r}-\tau _{e}\in ds|\tau _{r}=N].
\end{equation}%
Then by conditioning we have the following relations
\begin{equation}
\mathbb{P}_{x}[\tau _{r}-\tau _{e}\in ds]=\sum_{n=1}^{\infty }u_{n}(x)\phi
_{n}(x,s)ds.
\end{equation}%
\begin{equation}
\mathbb{P}_{x}[\tau _{r}-\tau _{e}\in ds|\tau _{r}<\infty ]=\dfrac{%
\sum_{n=1}^{\infty }u_{n}(x)\phi _{n}(x,s)ds}{\sum_{n=1}^{\infty }u_{n}(x)}.
\end{equation}%
Moreover, with the Markov property at time $(n-1)N$, we see that
\begin{equation}
\mathbb{P}_{x}[\tau _{r}-\tau _{e}\in ds|\tau _{r}=nN]=\int_{D}^{\infty }%
\mathbb{P}_{u}[\tau _{r}-\tau _{e}\in ds|\tau _{r}=N]\mathbb{P}_{x}%
[S_{(n-1)N}\in du|\tau _{r}=nN].  \label{eqn23}
\end{equation}%

This suggests that the distribution of $\tau _{e}-\tau _{r}$ is a mixture of
the distribution $\phi $. That is,

\begin{prop}
\label{prop3.1} Given a geometric L\'{e}vy process $(S_{t},t\geq 0)$ as in
expression (\ref{eqn:model}), the distribution of the time between economic
and recorded default is given by
\begin{equation}
\mathbb{P}_{x}[\tau _{e}-\tau _{r}\in ds]=\int_{D}^{\infty
}\sum_{n=1}^{\infty }\phi (u,s)u_{n}(x)\mathbb{P}_{x}[S_{(n-1)N}\in du|\tau
_{r}=nN].
\end{equation}
\end{prop}

Note that the above proposition in fact holds for any time homogeneous
stochastic processes with the strong Markov property.

When $(S_{t},t\geq 0)$ is a geometric spectrally positive Levy process, that
is,  $S_t=exp(X_t)$ with $X=(X_t)_{t\ge 0}$ a spectrally positive L\'{e}vy
process, the distribution of $\phi $ can be more explicitly computed.
Recall here that a spectrally positive L\'{e}vy process $X$ can be
represented as:
\begin{equation*}
X_{t}=b t+\sigma W_{t}+J_{t}^{+},
\end{equation*}%
where $W_{t}$ is a standard Brownian motion and $J^{+}=\{J_{t}^{+}\}$ is a
non-Gaussian L\'{e}vy process with positive jumps. 

\begin{prop}
\label{prop3.2} Assume that $S=(S_{t}, t \geq 0)$ is a geometric spectrally
positive L\'evy process. Let $S^{\prime }=(S^{\prime }_{t}, t \geq 0)$ be an
independent copy of $S$ with $S^{\prime }(0)=0$, then
\begin{equation}
\phi(x,s) =  \mathbb{P}_{x}[\tau_r - \tau_{e} \in ds | \tau_r = N] = \mathbb{P}[ \mathcal{A}_{H_{D}}
1_{H_{D} \leq N} \in ds].
\end{equation}
Here
\begin{equation}
H_{D} = \inf \{t; S_{t} \leq D \} \ \ and \ \ \mathcal{A}_{u} = \inf \{t
\leq N-u, \ S^{^{\prime }}_{t} \leq 0 | S^{^{\prime }}_{N-u} \leq 0 \}.
\end{equation}
\end{prop}

Proof: First, we see that
\begin{equation*}
\phi (x,s)= \mathbb{P}_{x}[\tau _{r}-\tau _{e}\in ds|\tau _{r}=N] =\mathbb{P}_{x}[\tau _{r}-\tau _{e}\in
ds|S_{N}\leq D],
\end{equation*}%
because $N$ is the first repayment date.

Therefore conditioned on the first hitting time of $H_{D}$, we get,
\begin{equation*}
\mathbb{P}_{x}[\tau _{r}-\tau _{e}\in ds|S_{N}\leq D]=\int_{0}^{N}\mathbb{P}%
_{x}[\tau _{r}-\tau _{e}\in ds|H_{D}\in du;S_{N}\leq D]\mathbb{P}%
_{x}[H_{D}\in du]
\end{equation*}%
Now, by the strong Markov property at the stopping time $H_{D}$ and $X$
being spectrally positive,
\begin{equation*}
\mathbb{P}_{x}[\tau _{r}-\tau _{e}\in ds|H_{D}\in du;S_{N}\leq D]=\mathbb{P}%
_{(u,D)}[\tau _{r}-\tau _{e}\in ds|S_{N}\leq D],
\end{equation*}%
where $\mathbb{P}_{(u,D)}$ is the distribution of $S$ starting from $D$ at
time $t=u$. This completes the proof.

\begin{rem}
Let $S=(S_{t},t\geq 0)$ be a geometric Brownian motion, i.e. $S_{t}=\exp
(\mu W_{t}-\mu ^{2}t/2)$ under the risk neutral measure for some constant $%
\mu >0$ with $\{W_{t}\}_{t\geq 0}$ a standard Brownian motion. Then, 
according to Bentata and Yor (2008) or Madan, Roynette, and Yor (2008),
\begin{equation}
\mathbb{P}_{(u,D)}[\tau _{e}-\tau _{r}\in ds \  |  \tau_{r} = N]=\dfrac{ds}{\pi \sqrt{s(N-u-s)}}%
\phi (\mu /2\sqrt{N-u-s}).
\end{equation}%
with $\phi (\mu )=\int_{0}^{\infty }dte^{-t}\cosh (\mu \sqrt{2t})$.
\end{rem}

\subsection{The distribution of $\protect\tau_r$}

This section focuses on $(\tau _{r},S_{\tau _{r}})$, the joint distribution
of the recorded default time and the value of the firm's assets. Recall that
$(S_{t},t\geq 0)$ is the exponential of a L\'{e}vy process. Therefore
\begin{equation}
(Y_{n})_{n \geq 1} = (\log {S_{nN}}-\log {S_{(n-1)N}})_{n\geq 1}
\end{equation}
is a sequence of i.i.d. random variables with a common distribution $F=^{d}
\log(S_{N})$. 

 For example, when $\log S_{t}$ is a standard Brownian motion with volatility $\sigma$,
$F=N(0,\sigma N)$. If $S_{t}$ is the exponential of an $\alpha $- stable
process, then $F$ is an $\alpha $- stable law. 

We now review some relevant notation and a useful lemma for a random walk
from fluctuation theory.

For a given random walk $(\log S_{nN},n\geq 0)$, one considers $T_{n}^{+}$ the $n
$-th time where the random walk $\log S_{nN}$ hits its maximum, and $%
H_{n}^{+}=\log S_{T_{n}^{+}N}$, the value of the random walk at that time.
Consequently, one defines the increasing ladder time $T^{+}_{n}$ and ladder
height process $H^{+}$ as $T^{+}=(T_{n}^{+},n\geq 0)$ and $%
H^{+}=(H_{n}^{+},n\geq 0)$ where
\begin{equation}
H_{n}^{+}=\log S_{T_{n}^{+}N}, \ T_{n}^{+}=\min \{r>T_{n-1}^{+}:\log S_{rN}>H_{n}^{+}\}
\end{equation}%
with $T_{0}^{+}=0.$ Similarly one can define the descending ladder time $%
T^{-} =  (T_{n}^{-}, n \geq 1)$ and ladder height process $H^{-} = (H_{n}^{-}, n \geq 1)$ as the sequence of consecutive times when the random walk
hits its minimum and the values at these minimum.

To study the distribution of the first hitting time of a random walk, it is
convenient to consider the Green function of the ladder process of the random walk:
\begin{equation}
U^{+}(dx, i) = \sum_{n = 0}^{\infty} P_r(H_{i}^{+} \in dx, \ T_{i}^{+} = n),
\end{equation}
and
\begin{equation}
U^{-}(dx, i) = \sum_{n = 0}^{\infty} P_r(H_{i}^{-} \in dx, \ T_{i}^{-} = n).
\end{equation}
The following lemma concerning the first hitting time of a random walk uses
the Green function of a ladder process:


\begin{lem}
Let $x \geq 0$ and for all $n \geq 0$, $Z_n = \max \lbrace 0,
\log S_{0},..., \log S_{nN} \rbrace$, $\bar{\theta}_{n} = \max_{k \leq n} \lbrace \log S_{kN}
= Z_n \rbrace $ and $\sigma_{x} = \min_{n \geq 1} \lbrace \log S_{nN} > x
\rbrace $, then for all $u >0$, $y \in [0,x]$, $v \geq y$, $i,j \in N$,
{\small \begin{eqnarray*}
P_r(\sigma_{x}-1-\overline{{\theta}_{\sigma_{x-1}}}=i,  \bar{\theta}%
_{\sigma_{x-1}} = j, &&  \log S_{\sigma_{x}N} -x \in du, x - \log S_{(\sigma_{x} -1)N} \in dv,
x - Z_{\sigma_{x} -1} \in dy ) \\
 = U^{+}(x-dy,j) &&  U^{-}(dv-y,i) F(du+v).
\end{eqnarray*}}
\end{lem}

Immediately, we have

\begin{thm}
\label{thm4.2} For any $y>0, k\ge 1$,
{\small \begin{equation}
P_x[\tau_r= kN,  \log S_{\tau_r } \in dy ] =  \sum_{i
= 0}^{k-1} \int_{\log(D)}^{\log(x)} U^{-}(\log(x/D) - dz,i) \int_{z}^{\infty} U^{+}(du-z,k-i) F( \log(D) -dy - u).
\end{equation}
}
\end{thm}

From a computational perspective, the Green functions can be derived by
inverting appropriate Laplace transforms. The following Friedst formula (see
\cite{Kyprianou2006} for more details) provides an analytical relation
between the Green functions of the ladder processes and the distribution of
the increments of the random walk.

\begin{thm}
(Friedst)
For all $(r,t) \in \mathbb{R}^{2}, $
\begin{equation}
1 - \mathbb{E}[r^{T_{1}^{+}} e^{itH_{1}^{+}}] = \exp( - \sum_{n=1}^{\infty}
\dfrac{r^{n}}{n} \mathbb{E}[e^{it \log (S_{Nn})} : S_{nN} > 1] ),
\end{equation}
\end{thm}

Finally, combining the previous two results, we get the distribution for $%
\tau _{e}$.

\begin{cor}
\begin{eqnarray*}
P_{r}(\tau _{e}\in du) &&=\sum_{k=1}^{\infty }P_{r}[\tau _{e}\in du|\tau
_{r}=kN]P_{r}[\tau _{r}=kN] \\
&=&\sum_{k=1}^{\infty }P[\tau _{r}-\tau _{e}\in d(kN-u)|\tau
_{r}=kN]P_{r}[\tau _{r}=kN] \\
&=&\sum_{k=1}^{\infty }P[\tau _{r}-\tau _{e}\in d(kN-u)|\tau
_{r}=kN]\int_{0}^{\infty }P[\tau _{r}=kN,S_{\tau _{r}}\in dy]
\end{eqnarray*}%
where $P[\tau _{r}=kN,S_{\tau _{r}}\in dy]$ follows from  Theorem \ref{thm4.2}%
, and $P[\tau _{r}-\tau _{e}\in ds|\tau _{r}=kN]$ is given by Equation (\ref%
{eqn23}) and Proposition \ref{prop3.2}.
\end{cor}

\subsection{Example}

To illustrate the above results, we provide two examples.

\begin{ex}
\label{ex1} In this example, we assume that firm's value $S$ is a geometric
Brownian motion with $\sigma =0.25,\mu =0.04$. We normalize the initial
value of the firm to be unity ($S_{0}=1$). The leverage ratio (debt to firm
value) is $0.8$, and the time between debt repayments is $N=15$ days with $%
D=0.4$.
\end{ex}

\begin{ex}
\label{ex2} In this example, we again assume that the dynamic of the firm's
value is a geometric Brownian motion with $\sigma =0.25,\mu =0.04$. The
leverage ratio is 0.8 and $N=3$ months with $D=0.1.$
\end{ex}

\begin{figure}[tbh]
\centering
\includegraphics[width=9cm]{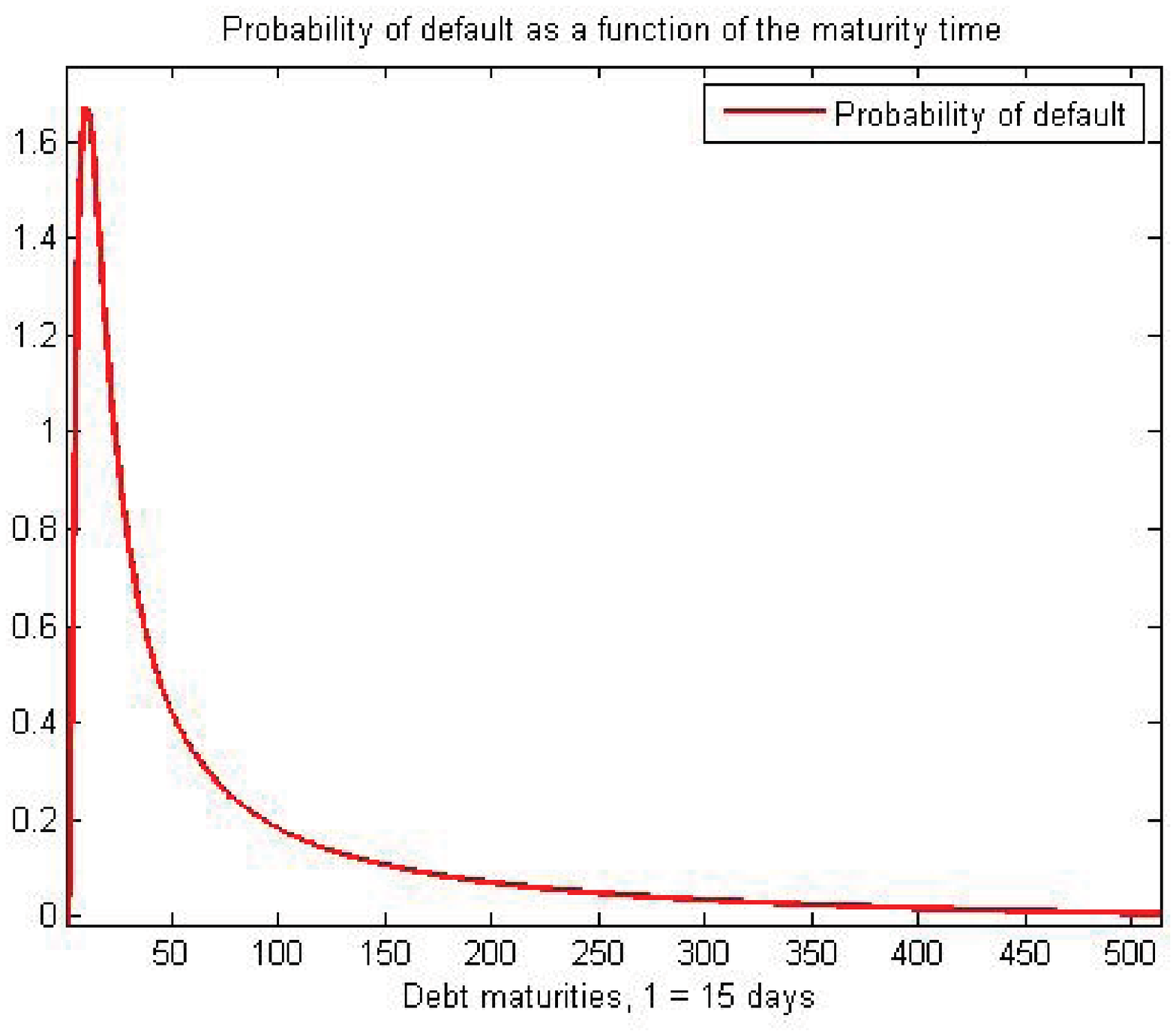}
\caption{Example \protect\ref{ex1}}
\label{figure2}
\end{figure}

\begin{figure}[tbh]
\centering
\includegraphics[width=9cm]{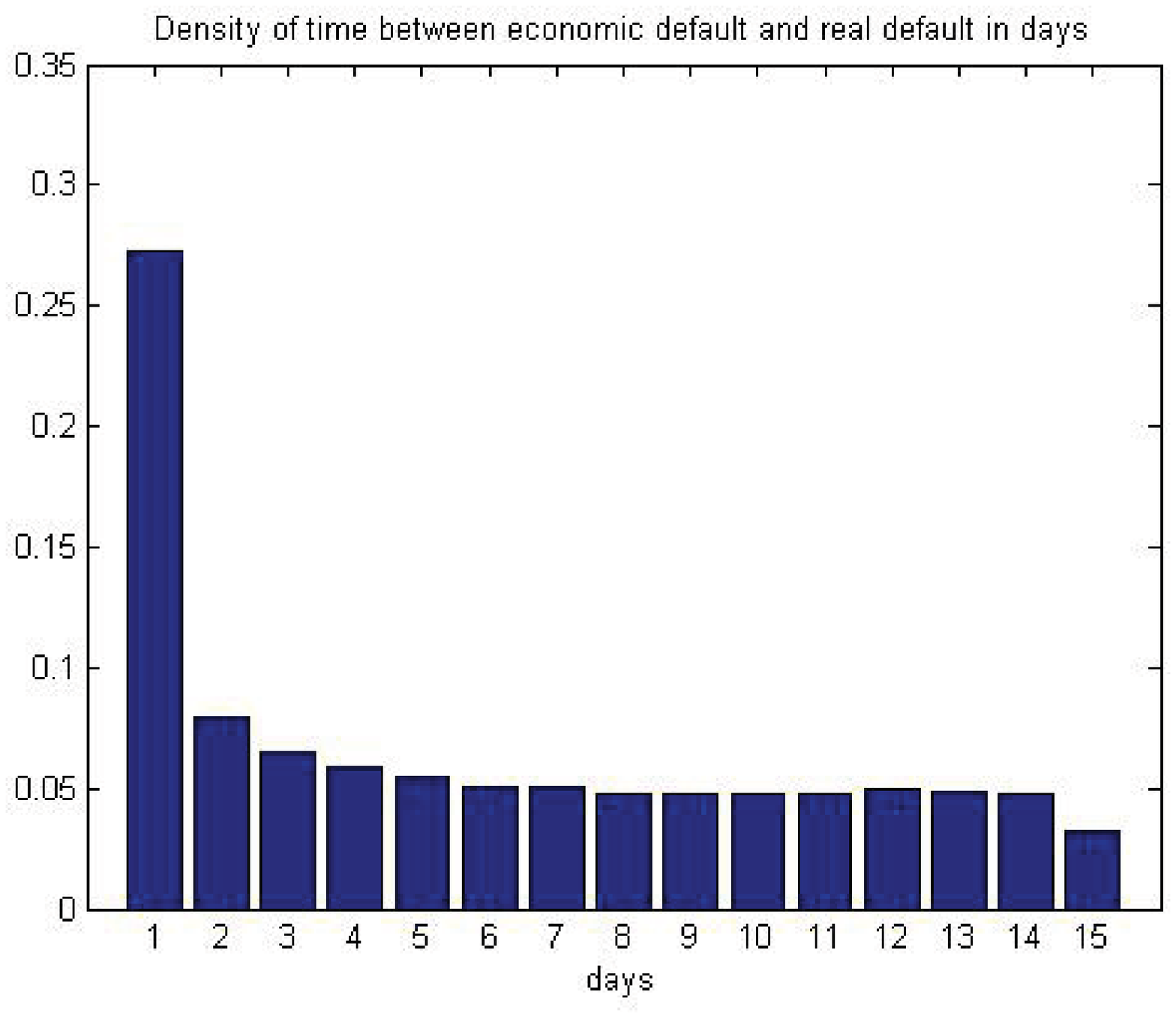}
\caption{Example \protect\ref{ex1}}
\label{figure3}
\end{figure}

\begin{figure}[tbh]
\centering
\includegraphics[width=9cm]{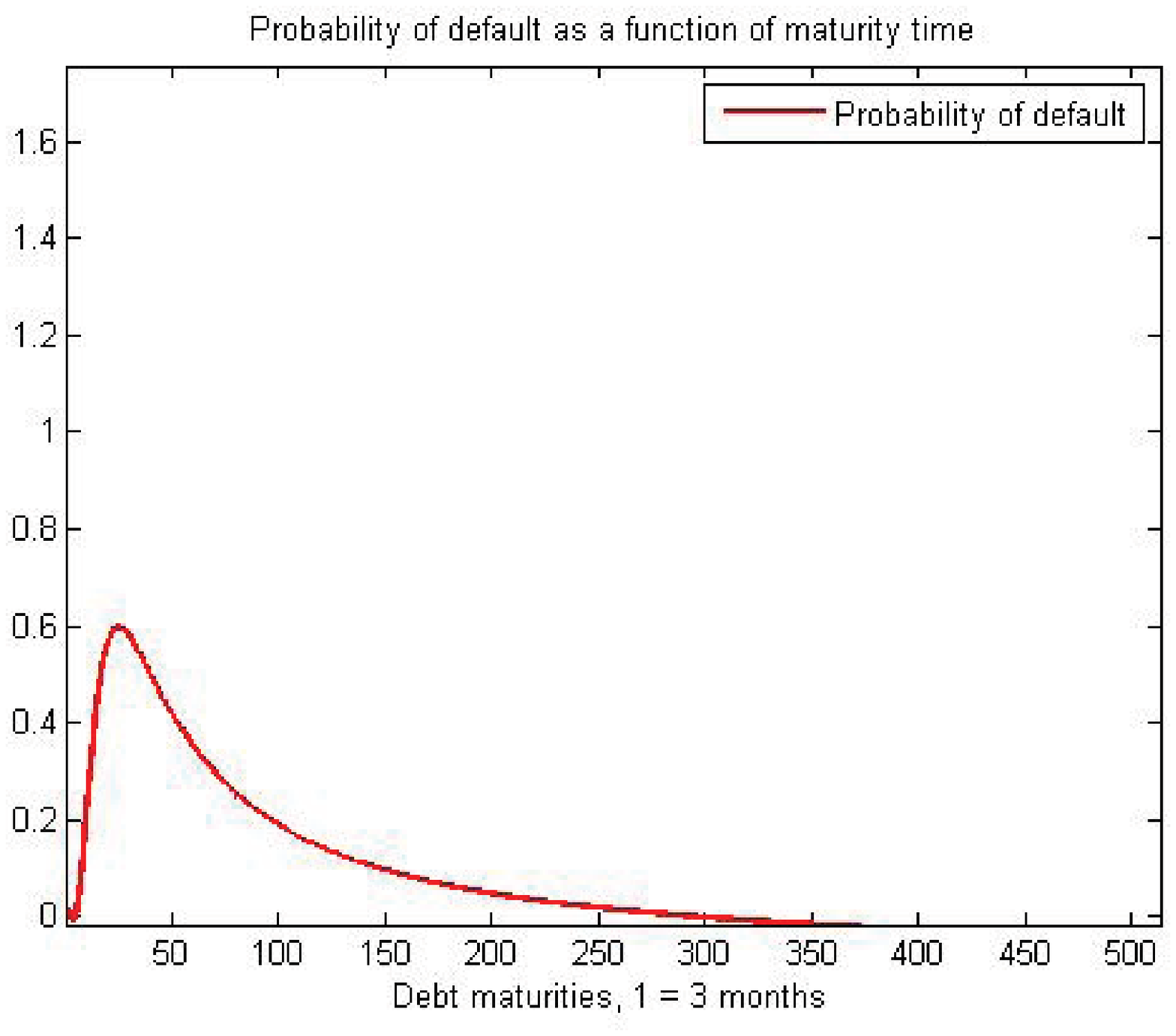}
\caption{Example \protect\ref{ex2}}
\label{figure4}
\end{figure}

\begin{figure}[tbh]
\centering
\includegraphics[width=9cm]{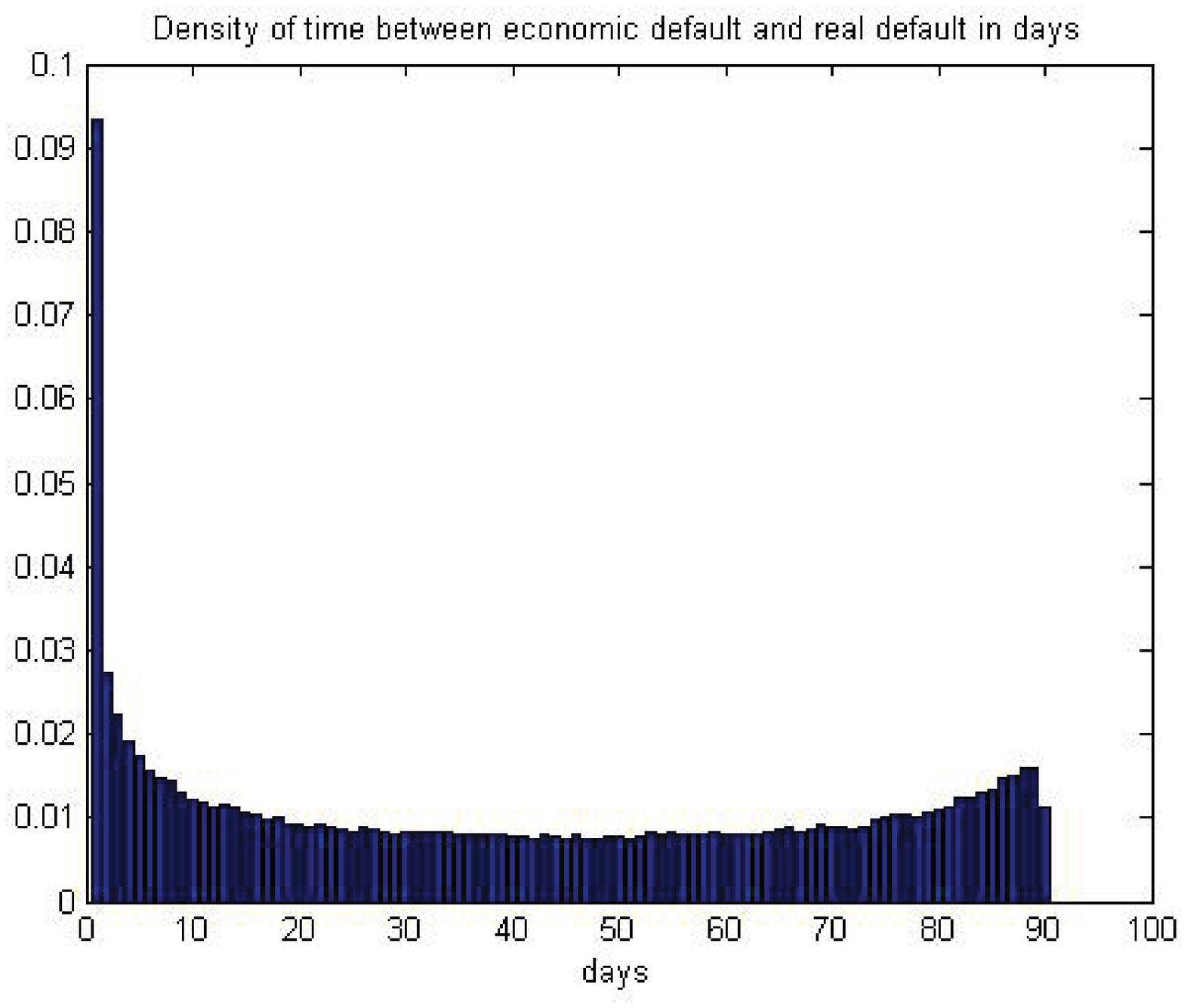}
\caption{Example \protect\ref{ex2}}
\label{figure5}
\end{figure}

\clearpage

\section{Relation with the Classical Structural Models}

In the classical structural models, the recorded default time is modeled as
the first hitting time of a diffusion process (with possible jumps) to a
default barrier in continuous time. In this section, we show that when
appropriately re-scaling the time step, our discrete model and the recorded
default time asymptotically approaches the classical definition of the
recorded default when the time step goes to zero. A useful corollary of our
approach is that one can use the tail index and correlation structure of $%
Y_{i}$ to estimate the parameters of the firm value process (either Brownian
motion based or more general L\'{e}vy models).

To start, consider a firm which reveals information only at the discrete
times $Nk$ for $k\geq 0$. The information is the current asset value $S_{Nk}$%
. Define for all $k\geq 1$, $\log (\dfrac{S_{Nk}}{S_{N(k-1)}})=Y_{k}$,
\textquotedblleft the firm's return\textquotedblright , with the sequence $%
(Y_{k},k\geq 1)$ being stationary. The debt is refinanced and rolls over at
these times $Nk$ and without loss of generality $r=0$.

Next, define $P_{x}^{N}(T_{1},T_{2})$ to be the probability that recorded
default will occur between $T_{1}$ and $T_{2}$ for any given $T_{2}>T_{1}>0$%
, i.e.,
\begin{equation}
P_{x}^{N}(T_{1},T_{2})=\mathbb{P}_{x}[T_{1}<\tau _{r}\leq T_{2}].
\end{equation}

Now consider the following re-scaling:

\begin{itemize}
\item re-scaling $N$ by a factor of $n$ with $N \mapsto N/n$, and

\item re-scaling the firm's return by a factor $\zeta _{n}$ (to be
determined) with $Y_{k}\mapsto \dfrac{Y_{k}}{\zeta _{n}}$,
\end{itemize}

so that for all $n\geq 1$,
\begin{equation}
S_{t}^{n}=S_{0}^{n}\exp (\sum_{k=1}^{[nt/N]}\dfrac{Y_{k}}{\zeta _{n}}).
\end{equation}%
Under the risk neutral measure, $S^{n}$ becomes
\begin{equation}
S_{t}^{n}=S_{0}^{n}\exp (\sum_{k=1}^{[nt/N]}\dfrac{Y_{k}}{\zeta _{n}}-\phi
_{n}(t)),
\end{equation}%
where for all $n\geq 1$,
\begin{equation}
\phi _{n}(t)=\log (\ \mathbb{E}[\exp (\ \sum_{k=1}^{[nt/N]}\dfrac{Y_{k}}{%
\zeta _{n}})]).
\end{equation}%
A simple application of the invariance principle will show that this
rescaling function $(\zeta _{n},n\geq 1)$ is a function of the tail exponent
of the return sequence $(Y_{k},k\geq 1)$.

\begin{prop}
If the sequence $(Y_{n}, n \geq 1)$ is a sequence of weakly dependent (such
as uniformly) stationary random variables with
\begin{equation}
\mathbb{E}[Y_{1}^{2}] + 2 \sum_{k=2}^{\infty} Cov(Y_{1},Y_{i}) = \sigma^{2}
< \infty,
\end{equation}
then $\zeta_{n} = \sqrt{n}$ and
\begin{equation}
(S^{n}_{t}, t \geq 0) \Rightarrow \exp( \dfrac{\sigma}{\sqrt{N}%
} B_{t} - t \dfrac{\sigma^{2}}{2N}) ,
\end{equation}
where the convergence is in distribution on the Skorokhod space $\mathcal{D}$%
. Moreover for all $T_{1} < T_{2}$,
\begin{equation}
P_{x}^{N/n}(T_{1},T_{2}) \rightarrow_{n \to \infty} \int_{T_{1}}^{T_{2}}
p(x,s,\dfrac{\sigma}{\sqrt{N}}) ds,
\end{equation}
for all $0 \leq s \le \infty$, with  $p(x,s,\dfrac{\sigma}{\sqrt{N}})$ the
first hitting time probability density function of $\exp( \dfrac{\sigma}{\sqrt{N}}
B_{t} - t \dfrac{\sigma^{2}}{2N})$ starting from x hitting the level $D$.
\end{prop}

\textbf{Proof.} First, let $\mathbb{E}[Y_{i}] = \mu$, one has from theorem
4.4.1 from \cite{Whitt2002}
\begin{equation}
(\dfrac{Y_{1} + ... + Y_{[nt]} - [nt] \mu }{\sqrt{n}} , t \geq 0 )
\Rightarrow ( \sigma B_{t}, t \geq 0 ) \ \ \mbox{on} \ (\mathcal{D}, J_1).
\end{equation}
Therefore, by continuity of the exponential function on the Skorokhod space,
one has
\begin{equation}
(\exp(\sum_{i=0}^{[nt/N]} \dfrac{Y_{i}}{\sqrt{n}} - t \phi_{n}) , t \geq 0)
\Rightarrow \exp( \dfrac{\sigma}{\sqrt{N}} B_{t} - t \dfrac{%
\sigma^{2}}{2N}) \ \ \mbox{on} \ (\mathcal{D}, J_1).
\end{equation}
Moreover, since the first hitting time is a continuous function on the
Skorokhod space,
\begin{equation}
P_{x}^{N/n}(T_{1},T_{2}) \rightarrow_{n \to \infty} \int_{T_{1}}^{T_{2}}
p(x,s,\dfrac{\sigma}{\sqrt{N}}) ds,
\end{equation}
where for all $0 \leq s \le \infty$, $p(x,s,\dfrac{\sigma}{\sqrt{N}})$ is
given by
\begin{equation}
p(x,s,\dfrac{\sigma}{\sqrt{N}}) = \dfrac{\log(D/x)}{\sqrt{2 \pi N}s^{3/2}
\sigma} \exp(- \dfrac{(\dfrac{\log(D/x)}{\sqrt{N} \sigma} + \dfrac{\sigma s}{%
2\sqrt{N}})^{2}}{2s} )
\end{equation}

\begin{rem}
When $N$ goes to zero, where the dynamics of the firm value follows a
geometric Brownian motion, it is clear $\tau _{r}-\tau _{e}=0$ almost surely.
\end{rem}

When the sequence $(Y_{i},i\geq 1)$ does not have a moment of order two, we
see that

\begin{prop}
If
\begin{equation}
Y_{1}+...+Y_{n}=^{d}n^{1/\alpha }Y_{1},
\end{equation}%
for $\alpha \in (0,2)$ then $\zeta_{n} = n^{1/\alpha}$ and
\begin{equation}
S^{n} \Rightarrow \exp(S^{\alpha}_{t} - t C_{\alpha}) \ \ \mbox{on} \ (\mathcal{D}, J_1),
\end{equation}
where $C_{\alpha}$ is a constant depending on the characteristics of the
stable process $S^{\alpha}$, with $\tau_r=\tau_e=\inf\{t, S_t\le D\}$.
\footnote{%
This is known as $\alpha $-stable distribution, which is generally
represented as a four-parameter family: the index, the scale, the skewness
and the shift parameter. For more details, see for instance \cite{ST1994}
and \cite{Whitt2002}).}
\end{prop}

\begin{rem}
The previous proposition links the volatility coefficient of the firm value
process $\sigma _{f}$ as a function of the variance of its return $Y_{i}$
and $N$ so that
\begin{equation}
\sigma _{f}^{2}=\dfrac{\mathbb{E}[Y_{1}^{2}]+2\sum_{k=2}^{\infty
}Cov(Y_{1},Y_{k})}{N}.
\end{equation}%
Thus, one can use the tail index of $Y_{1}$ and the correlation structure of
the sequence $(Y_{i},i\geq 1)$ to estimate the parameters of the firm's
value process.
\end{rem}

\section{Conclusion}

This paper develops a structural credit risk model to characterize the
difference between the economic and recorded default times for a firm.
Recorded default occurs when default is recorded in the legal system. The
economic default time is when the market first prices the firm's debt as if
it has defaulted. It has been empirically documented that these two times
are distinct (see Guo, Jarrow, and Lin (2008)). In our model, this
distinction is obtained by assuming that debt repayments occur on a set of
discrete dates, and not continuously as in the classical structural models.
We show that the probability distribution for the time span between economic
and recorded defaults follows a mixture of Arcsine Law, which is consistent
with the results contained in Guo, Jarrow, and Lin. In addition, we show
that the classical structural model can be obtained as a limiting case of
our model as the time between the debt repayment dates goes to zero. We also
illustrate how to estimate the parameters of the firm value process using
the time series of the firm's return. Our model is a first step in
characterizing the difference between the economic and recorded default
times. Our simple model can be generalized in numerous ways, and we hope to
pursue these extensions in subsequent research.

\end{document}